\documentstyle[preprint,pre,aps]{revtex}
\newcommand{\beq}{\begin{equation}}
\newcommand{\eeq}{\end{equation}}
\begin{document}
\draft
\preprint{Preprint Version 4.0, March 2002}
\title{Magneto-chiral scattering of light: a new optical manifestation
of chirality}
\author{ 
${\mbox{F.\,A.\,Pinheiro}}$~\cite{Mail}
${\mbox{and B.\,A.\,van Tiggelen}}$
}
\address{CNRS/Laboratoire~de~Physique~et~Mod\'elisation~des~Milieux~Condens\'es,
Universit\'e~Joseph~Fourier,~Maison~des~Magist\`eres,\\
B.P. 166~38042~Grenoble~Cedex~9,~France.\\
}
\date{\today}
\maketitle
%
%
%%%%%%%%%%%%%%%%%%%%%%%%%%%%%%%%%%%%%%%%%%%%%%%%%%%%
%%%
%
% A B S T R A C T:
%
%%%%%%%%%%%%%%%%%%%%%%%%%%%%%%%%%%%%%%%%%%%%%%%%%%%%
%%%
%
%
\begin{abstract}

	We have investigated multiple scattering of light in systems subject to
magneto-chiral (MC) effects. Our medium consists of magneto-optically active
dipoles placed in a
chiral geometry under the influence of an external
magnetic field. We have calculated, for the first time, the total and the differential
scattering MC cross sections of this system, explicitely showing that they are
proportional to {\em pseudoscalar} quantities. This
provides a new optical measure for the degree of chirality, a pseudoscalar $g$,
of an arbitrary geometrical configuration of scatterers based on its scattering properties.
We have calculated $g$ for some simple
chiral systems and we have even used it to probe the degree of optical chirality of random systems.
Finally, we have compared $g$ with other recently defined chiral measures in
literature.

\end{abstract}
+%
%%%%%%%%%%%%%%%%%%%%%%%%%%%%%%%%%%%%%%%%%%%%%%%%%%%%
%%%
%
% T E X T   O F   P A P E R:
%
%%%%%%%%%%%%%%%%%%%%%%%%%%%%%%%%%%%%%%%%%%%%%%%%%%%%
%%%
%
%
%
%
%%%%%%%%%%%%%%%%%%%%
% Section 1:
%%%%%%%%%%%%%%%%%%%%
%
%
\narrowtext
\section{Introduction}
\label{sec:intro}

	The breaking of fundamental symmetries of nature manifests itself in various
new optical phenomena and is essential for our understanding of the
interaction between light and complex matter. One well-known example is
the natural
optical activity discovered by Arago in 1811, a nonlocal optical
response in media that have broken mirror symmetry, named {\em chiral} media.
Another example is the magnetically induced optical activity, first
observed by Faraday in 1846~\cite{faraday} and associated with
the breaking of time-reversal symmetry by a
magnetic field. Although physically distinct, the two effects 
manifest themselves quite similarly in homogeneous media as a rotation of
the linear polarization of light. This resemblance has motivated
numerous works, pioneered by Pasteur~\cite{pasteur}, to search for a link between
optical activity and the Faraday effect, but in vain. The existence of
an optical effect 
providing such a link is only allowed to occur under conditions where both
mirror and time-reversal symmetries are simultaneously broken. The
cross-effect between natural and magneto optical activity is called the  
{\em magneto-chiral} (MC) {\em effect} and was first predicted in
1962~\cite{groenewege}, followed
by studies in crystalline materials~\cite{portigal}. The MC effect was later again
predicted independently several times~\cite{baranova77,baranova79,wagniere82,wagniere84},
both for absorption and for refraction.
 
	In order to deduce the MC effect in homogeneous and isotropic media, let us consider the
propagation of light with wavevector $\bf{k}$ through a magneto-optically active chiral
medium with dielectric tensor $\mathbf{\varepsilon}$ under the influence of a
static magnetic field ${\bf B}$. Expanding $\mathbf{\varepsilon }$ to first
order in ${\bf k}$ and ${\bf B}$, we have~\cite{portigal}:

\begin{equation}
\label{eq:epsilon}
\varepsilon _{ij}(\omega ,{\bf k}, {\bf B})=\varepsilon (\omega ) \delta_{ij} + \alpha(\omega) i \epsilon
_{ijl}k_{l} + \beta(\omega) i \epsilon _{ijl}B_{l} + \gamma(\omega) ({\mathbf B \cdot k})  \delta_{ij}
\end{equation}
where $\epsilon_{ijk}$ is the Levi-Civita tensor, $\omega /2 \pi$ is the optical frequency and $\alpha$ and $\beta$
are associated with the natural and magneto-optical activity, 
respectively. The last term of Eq.~(1) describes the
MC effect in homogeneous and isotropic media. It is important to point out some of the major
features related to this effect. Firstly, it depends on the
relative orientation of {\bf k} and {\bf B}. Secondly, the effect has opposite sign for the
two different chiral enantiometers. Finally, it is independent of the state of
polarization of light. Aside from its  
fundamental value for the understanding of the interaction between light and matter
and the underlying symmetry principles that govern it, the
MC effect in absorption has been suggested to be at
the origin of the homochirality of life~\cite{wagniere83}, since it allows
enantioselective photochemistry in a magnetic field
with unpolarized light~\cite{rikken2000}.   
Unfortunately, the magnitude of the molecular MC effect is very weak (typically of the order of $10^{-6}$)
since it can be regarded, as discussed above, as a cross-effect between two
already small effects, the natural and magneto optical activity.
That's why the MC effect has only been reported recently:
in absorption by Rikken and Raupach~\cite{rikken97}, and in refraction
by Kleindienst {\sl et al.}~\cite{kleindienst} 
and Vallet {\sl et al.}~\cite{vallet}.
Also the electrical analogue of the MC effect was
recently reported~\cite{rikken2001}.

	In the present paper, we study the MC effect in 
{\em scattering} of light. We have
investigated {\em multiple} light scattering by {\em magneto-optically active}
scatterers distributed in a {\em chiral} configuration under the influence of an
external magnetic field. This approach is completely
different from the traditional molecular one that has been employed to model the MC effect~\cite{baranova77,baranova79,wagniere82,wagniere84}.
Multiple light scattering has been studied for scatterers exhibiting rotatory power~\cite{ablitt}.
In our model magneto-chirality is a {\em collective} effect
built up by multiple scattering.
One single scatterer does not exhibit MC effects since it
is not made of a material that is natural optically active. However, an {\em assembly} of
magneto-optically active particles distributed in a {\em chiral} configuration
should generate MC effects since in such a system
both mirror and time-reversal symmetry are broken.  

	In order to investigate how scattering is affected by the MC effect,
we have calculated, to our knowledge for the first time, the differential
and the total cross-sections for a system composed by an arbitrary
number of magneto-optically active scatterers in a magnetic field. We have investigated if
these quantities are sensitive to the degree of chirality associated with the geometrical
configuration of the scatterers in space, and have concluded that scattering
in MC media constitutes a new optical manisfestation of chirality,
in addition to the well-known rotatory power.
A new optical parameter is introduced to quantify the chirality
associated with the spatial geometry, which we have
calculated for some simple chiral systems. We have also employed this parameter
to probe the chirality of random systems, which are in general chiral.
Finally, we have compared it to recently
defined chiral measures in literature~\cite{harrisprl,harrisrmp,harrisferro}.

	This paper is organized as follows. In Sec. II we briefly
explain how to quantify chirality and we explore this concept to define
a new optical chiral parameter based on the scattering properties of
MC media. In Sec. III we calculate this parameter for the simplest chiral
geometry, the so-called ``twisted H", whereas in Sec. IV we employ it to
quantify the degree of chirality of random scattering systems. Finally,
in Sec. V we summarize the main results of this paper.
 
%
%%%%%%%%%%%%%%%%%%%%
% Section 2:
%%%%%%%%%%%%%%%%%%%%
%

\section{Multiple scattering in magneto-chiral media}
\label{sec:multiple}
	There are several scattering systems eligible for MC effects.
The most obvious of them could be
a system composed by an assembly of MC scatterers (i.e., made of
a material exhibiting both the Faraday effect and natural optical activity).
In order to study such a system,
one should know precisely how an individual MC object scatters light. 
Mathematically, this means that the scattering $T$ matrix of this object
should be known. However, to our knowledge, no
calculation exists for such a $T$ matrix. 
Alternatively, it is possible to
generate the MC effect in scattering systems
by considering magneto-optically active scatterers
distributed in chiral geometries.
This is the kind 
of system that
will be treated in the present work. Before we describe the
light scattering properties of such system,
it is instructive to make some general 
considerations about chirality itself
and how one can quantify the degree of chirality associated with
an arbitrary object or a distribution of scatterers in space.
Afterwards, we will discuss how a chiral configuration of scatterers
affects the properties of the light scattered in MC media. 

\subsection{What is chirality and how to quantify it?}
\label{sec:chiral}

By definition, a chiral object is 
an object whose mirror image cannot be rotated to coincide
with itself.
It can be readily seen that this implies to have at least four particles
for the system to be chiral. A three-particle system 
is necessarily achiral since it is always contained
in a plane, which is automatically its mirror plane. 

	Recent work has addressed the problem whether chirality
can be {\em quantified} or {\em measured},
in as much the same way that one can quantify the degree of order of a ferromagnet.
Despite the ubiquity of chirality in nature (almost all arbitrary microscopic and macroscopic 
objects are chiral, from DNA to a piece of rock) and the important
role played by chirality in several areas of research, from pharmaceutics
to liquid crystals~\cite{gennes} and to the origin of life~\cite{wagniere83},
only recently real progress has been made.
One of the first attempts to establish a quantitative measure of
chirality is due to Gilat~\cite{gilat}.
Osipov {\em et al.}~\cite{osipov} have developed a molecular measure
of chirality based on the behaviour of the response functions
that characterize molecular optical activity.
Harris {\em et al.}~\cite{harrisprl,harrisrmp,harrisferro}
have proposed a new elegant way, based on group theory, to measure the degree of chirality
exhibited by a geometric object. They
have constructed a variety of rotationally invariant pseudoscalars
and have applied them
to quantify the degree of chirality of molecules of arbitrary shape, showing how these
parameters govern a particular observable, such as the pitch of a
cholesteric liquid crystal~\cite{harrisprl,harrisrmp,harrisferro}
and the optical rotatory power~\cite{spector2000}. 
	
	The choice for a chiral ``order" parameter is not unique~\cite{harrisrmp}, but
must always be
a pseudoscalar {\em invariant under rotations}.
This last fact guarantees that no rotation of the object exists
that maps the mirror image of a chiral object onto itself.

	In what follows, we will describe the scattering properties
of an assembly of magneto-optically active dipoles in a magnetic field.

\subsection{Scattering by magneto-optically active particles}
\label{sec:scattering}

	In our model, we deal with dielectric {\em pointlike} magneto-optical scatterers 
(i.e., radiative dipoles much
smaller than the wavelength of light) 
located in
vacuum. This model enables us to find exactly 
all scattering properties (i.e., the off-shell $T$ matrix) of any number of scatterers
in analytic form and without any further
approximation. This model exhibits a scattering resonance, where the scattering
cross-section of a 
single scattering reaches its maximum value.
The $3 \times 3$ $T$ matrix for a single scattering
takes the form~\cite{nieuwen96}:

\begin{equation}
\label{eq:tmatrix}
{\bf t}({\bf B},\omega )= (t_{0} - t_{3}) {\bf U} + t_{1} {\bf \Phi }
+t_{2} \bf{BB},  
\end{equation}
where ${\bf U}$ is the unitary matrix, $t_{0}$ is the ordinary Rayleigh $T$ matrix,
$\mathbf{\Phi }$ is the
antisymmetric Hermitian tensor $\Phi _{ij}=i \epsilon _{ijk}B_{k}$ and the complex-valued parameters $t_{1},t_{2},t_{3}$
are defined in Ref.~\cite{nieuwen96}.
	The knowledge of the single scattering $T$ matrix in Eq.~(2)
allows us to obtain the total $3N \times 3N$ $T$ matrix of
an assembly of $N$ magneto-optical
scatterers situated at the positions ${\bf r}_{1}, {\bf r}_{2},...,{\bf r}_{N}$:

\begin{equation}
\label{eq:totmatrix}
{\bf T}_{{\bf k,k^{'}}} = \left( 
\begin{array}{c}
e^{i {\mathbf k\cdot r}_{1}} \\ 
: \\ 
e^{i {\mathbf k\cdot r}_{N}}
\end{array}
\right) ^{\ast }{\mathbf t} {\mathbf \cdot} ({\mathbf U} - {\mathbf G} {\mathbf \cdot} {\mathbf t})^{-1} {\mathbf \cdot} \left( 
\begin{array}{c}
e^{i{\mathbf k}^{\prime }{\mathbf \cdot r}_{1}} \\ 
: \\ 
e^{i{\mathbf k}^{\prime }{\mathbf \cdot r}_{N}}
\end{array}
\right),
\end{equation}
where ${\bf k}$ and ${\bf k^{'}}$ are,
respectively, the incident and the scattered wave vectors,
and $|{\mathbf k}^{\prime }| = |{\mathbf k}| = k$.
The elements of the $3N \times 3N$ $\bf{G}$-matrix are equal to
the Green functions calculated from the relative
positions of the scatterers~\cite{reports}:

\begin{equation}
\label{eq:green}
{\mathbf G}_{NM}=\left\{ 
\begin{array}{c}
-\frac{\exp (ikr_{_{NM}})}{4\pi r_{_{NM}}}\text{\ }\left\{ \left[ {\mathbf U}-
\widehat{{\mathbf r}}_{_{NM}} \widehat{{\mathbf r}}_{_{NM}}\right] -\left( 
\frac{1}{ikr_{_{NM}}}+\frac{1}{(kr_{_{NM}})^{2}}\right) \left[ {\mathbf U}-3
\widehat{{\mathbf r}}_{_{NM}}\widehat{{\mathbf r}}_{_{NM}}\right] \right\} 
\text{\ \ \ \ for \ \ }N\neq M, \\ 
0\ \ \ \ \ \ \ \ \ \ \ \ \ \ \ \ \ \ \ \ \ \ \ \ \ \ \ \ \ \ \ \ \ \ \ \ \ \ \ \ \ \ \ \ \ \ \ \ \ \ \ \ \ \ \ \ \ \ \ \ \ \ \ \ \ \
\ \ \ \ \ \ \ \ \ \ \ \ \ \ \ \ \ \ \ \ \ \ \ \
\text{for \ \ }N=M.
\end{array}
\right. 
\end{equation}
The set of Eqs.~(3) form a system of linear equations that determines
the electric field scattered by the magneto-optical particles, and its
solution requires the diagonalization
of the $3N \times 3N$ scattering
matrix~\cite{polonia}:

\begin{equation}
\label{eq:scatteringmatrix}
{\mathbf M} (k) \equiv {\mathbf t}(k) {\mathbf \cdot} [{\mathbf U} - {\mathbf G}(k) {\mathbf \cdot} {\mathbf t}(k) ]^{-1}.
\end{equation}
We have modified the numerical code developed by Rusek and Orlowski~\cite{polonia} in order to
diagonalize the scattering matrix ${\mathbf M} (k)$ in Eq.~(5)
for Faraday-active scatterers. This provides the scattered field
amplitude for particles distributed in an arbitrary spatial configuration:
\begin{eqnarray}
\label{eq:field1}
\nonumber f (\sigma {\bf k} \rightarrow \sigma^{'} {\bf k^{'}})
= {\bf g}^{\ast}_{\sigma} \cdot {\mathbf f} ( {\bf k} \rightarrow {\bf k^{'}}) \cdot {\bf g}_{\sigma^{\prime}}
\text{ \ \ \ \ with }  \\
{\mathbf f} ( {\bf k} \rightarrow {\bf k^{'}}) = \sum_{NM} {\bf M}^{NM} 
\exp( i {\bf k} \cdot {\bf r}_{N}) \exp(- i {\bf k^{'}} \cdot {\bf r}_{M}),  
\end{eqnarray}    
where ${\bf g}_{\sigma}$ and ${\bf g}_{\sigma^{'}}$ are, respectively, the incident 
and scattered polarization vectors (expressed in the circular polarization basis) and the
sum is performed over all scatterers.
Since the Faraday effect is a small perturbation to the scattering, 
we will expand ${\mathbf f}$ {\em linearly}
in the magnetic field, ${\mathbf f} = {\mathbf f}^{0} + {\mathbf f}^{1}$,
with ${\mathbf f}^{1}$ given by:

\begin{equation}
\label{eq:field2}
{\mathbf f}^{1}({\bf k} \rightarrow {\bf k^{'}},{\mathbf B} ) = - \frac{k}{4 \pi} \alpha (B) \sum_{NMN^{'}} 
({\bf M}_{0}^{NM} {\mathbf \cdot} {\mathbf \Phi} {\mathbf \cdot} {\bf M}_{0}^{MN^{'}}) 
\exp( i {\bf k} \cdot {\bf r}_{N}) \exp(- i {\bf k^{'}} \cdot {\bf r}_{N^{'}}),  
\end{equation}
where $\alpha(B)$ is a dimensionless quantity that measures the strength of 
the magnetic field and ${\bf M}_{0}$ is the $3 \times 3$ scattering matrix for ${\mathbf B} = 0$.
For an atom subject to the Zeeman effect, $\alpha (B)$ is typically
the ratio between the Zeeman splitting and the resonance linewidth~\cite{miniat}.
The corresponding {\em magneto extinction cross section} $\sigma^{1}_{ext} ({\bf B},{\bf k})$,
linear in the magnetic field and integrated over
all scattered angles, can be obtained
from the optical theorem~\cite{newton}:

\begin{eqnarray}
\nonumber \sigma _{ext}^{1}({\mathbf B},{\mathbf k}) &=& -\frac{1}{k}
\text{Im} \sum_{\sigma } f^{1}(\sigma {\mathbf k}\rightarrow \sigma {\mathbf k})
\\&=&\frac{\alpha (B)}{4\pi }\text{Im}\sum_{NMN^{^{\prime }}} \text{Tr} 
\left( {\mathbf M}^{NM}_{0} \cdot {\mathbf \Phi }\cdot {\mathbf M}_{0}^{MN^{\prime
}}\cdot {\mathbf \Delta }_{\mathbf k} \right) \exp \left[ i {\mathbf k}\cdot \left( 
{\mathbf r}_{N}- {\mathbf r}_{N^{\prime }}\right) \right], 
\end{eqnarray}
where $({\bf \Delta_{k}})_{ij} 
\equiv \delta_{ij} - k_{i} k_{j}/k^{2}$ is the
projector upon the space of transverse polarization, normal to ${\bf k}$.
Notice that we have traced over polarization, which means that
the Faraday effect itself cancels and that only MC effects remain.
Eq.~(8) can be regarded as the magneto dichroism of the
multiple scattering system, associated with the energy removed from the incident beam
due the application of an external magnetic field. 

	Until now, nothing was said about how the light scattering
properties of magneto-optically active particles are affected
by the chiral configuration and, more important, 
how these properties can be related to a measure of the degree of chirality.
We recall
that any chiral measure must be rotationally invariant. This implies that we must
perform an {\em average} over
{\em all solid rotations} in
the expression (8) for $\sigma^{1}_{ext} ({\bf B},{\bf k})$
in order to estimate magneto-optical chirality.
The angularly averaged magneto extinction
cross-section in MC media must necessarily have the form:
\begin{equation}
\langle \sigma _{ext}^{1}({\mathbf B},{\mathbf k}) \rangle_{4 \pi} = 
g \alpha(B) ( \mathbf{\widehat{B}} \cdot \mathbf{\widehat{k}}),
\end{equation}
where $g$ is some pseudoscalar and the brackets $\langle...\rangle_{4 \pi}$
denote the average over all solid rotations.
This is mathematically equivalent to an average over the vectors
${\mathbf k}$ and
${\mathbf B}$ with {\em fixed} mutual orientation.  
Equation (9) is manifestly
rotationally invariant and obeys the two fundamental
symmetry relations, parity and reciprocity, in MC media, as can easily be verified.
Using Eqs. (8) and (9), we can write $g$ as: 

\begin{equation}
\label{sigmatotmed}
g \equiv \frac{1}{4\pi }\text{Im}\sum_{NMN^{^{\prime }}}  \left\langle \text{Tr}%
\left( {\mathbf M}_{0}^{NM} \cdot {\mathbf \Phi_{k} }\cdot {\mathbf M}_{0}^{MN^{\prime
}}\cdot {\mathbf \Delta }_{\mathbf k}\right) \exp \left( i {\mathbf k}\cdot  
{\mathbf r}_{N N^{\prime }} \right) \right\rangle_{ {\mathbf k} \in 4 \pi},
\end{equation}
where ${\mathbf r}_{N N^{\prime }} \equiv {\mathbf r}_{N^{\prime}} - {\mathbf r}_{N}$ and 
$(\Phi_{\mathbf{k}})_{ilm} = i \epsilon _{ilm} k_{m}$.
Notice that, since ${\bf M}^{NM}_{0}$ depend only on
the positions of the scatterers $\left\{ {\mathbf r}_{N} \right\}$ and not on
$\widehat{{\mathbf k}}$, they will not be affected by the average.
Consequently, the term into the brackets in Eq.~(10)
involves an average of the type
$\langle \widehat{{\mathbf k}} (1- \widehat{ {\mathbf k}} \widehat{{\mathbf k}} )
\exp ( i {\mathbf k}\cdot  
{\mathbf r} ) \rangle_{{\mathbf k} \in 4 \pi}$,
that can be obtained analytically as:

\begin{equation}
\label{average2}
\left\langle \widehat{k}_{i}\left( \delta _{jk}-\widehat{k}_{j}\widehat{k}
_{k}\right) \exp \left( i{ \mathbf k\cdot r}\right) \right\rangle _{{\mathbf k} \in 4 \pi}
= i\widehat{r}_{i}\delta _{jk}j_{1}(kr)   
-i\left[ \widehat{r}_{i}\delta _{jk}+\widehat{r}_{j}\delta _{ik}+\widehat{r
}_{k}\delta _{ij}\right] \frac{j_{2}(kr)}{kr}  
+ij_{3}(kr)\widehat{r}_{i}\widehat{r}_{j}\widehat{r}_{k},
\end{equation}
where $j_{n}(x)$ is the spherical Bessel function of the first kind and of order $n$.
In addition, since
$ (M_{0}^{NM})_{ij} = (M_{0}^{MN})_{ji} $, we have
$\text{Tr}({\mathbf M}_{0}^{NM} \cdot {\mathbf \Phi_{k} }\cdot {\mathbf M}_{0}^{MN^{\prime}}) = 
- \text{Tr}({\mathbf M}_{0}^{N^{\prime}M} \cdot {\mathbf \Phi_{k} }\cdot {\mathbf M}_{0}^{MN})$.
As a result, the terms with $N = N^{\prime}$ in the sum of Eq.~(10) vanish.
Consequently, we can simplify Eq. (10) to:

\begin{equation}
\label{sigmatotmed2}
g = \frac{1}{2 \pi }\text{Im}\sum_{N^{^{\prime }} < N,M} i \left \langle \text{Tr}%
\left( {\mathbf M}_{0}^{NM} \cdot {\mathbf \Phi_{k} }\cdot {\mathbf M}_{0}^{MN^{\prime
}}\cdot {\mathbf \Delta }_{\mathbf k}\right) \sin \left( {\mathbf k}\cdot  
{\mathbf r}_{N N^{\prime }} \right) \right \rangle_{{\mathbf k} \in 4 \pi}.
\end{equation}
This clearly represents
a pseudoscalar, since under a mirror operation $\left\{ {\mathbf r}_N \right\}$
and ${\mathbf M}_0 ( \left\{ {\mathbf r}_N \right\} )$ transform as 
${\mathbf r}_N \rightarrow - {\mathbf r}_N$ and
${\mathbf M}_0 (\left\{ {\mathbf r}_N \right\} ) \rightarrow {\mathbf M}_0 (\left\{ -{\mathbf r}_N
\right\} ) = {\mathbf M}_0 (\left\{ {\mathbf r}_N \right\} )$.
Inserting Eq.~(11) into Eq.~(12), it follows that:

\begin{eqnarray}
\nonumber g 
=\frac{1}{2 \pi }\text{Re} \sum_{N^{\prime} < N,M} \text{%
Tr}\left( {\mathbf M}_{0}^{NM}\cdot {\mathbf \Phi }_{\widehat{{\mathbf r}}%
_{NN^{\prime }}}\cdot {\mathbf M}_{0}^{MN^{\prime }}\right) \left[
j_{1}(kr_{NN^{\prime }})-\frac{j_{2}(kr_{NN^{\prime }})}{kr_{NN^{\prime }}}%
\right]   \\
+\left. \left( \widehat{{\mathbf r}}_{NN^{\prime }}\cdot {\mathbf M}_{0}^{NM} \cdot 
{\mathbf \Phi }_{\widehat{{\mathbf r}}_{NN^{\prime }}}\cdot {\mathbf M}_{0}%
^{MN^{\prime }}\cdot \widehat{{\mathbf r}}_{NN^{\prime }} \right)
j_{3}(kr_{NN^{\prime}}) \right. . 
\end{eqnarray}	
	
	The second observable of interest is the angular distribution
of the light scattered by all particles. With this purpose, we have calculated
the differential magneto cross section linear in ${\mathbf B}$,
$\frac{d \sigma^{1}}{d \Omega} (\sigma {\bf k} \rightarrow \sigma^{'} {\bf k^{'}}, {\mathbf B})$, using the relation:

\begin{equation}
\label{eq:dif1}
\frac{d \sigma}{d \Omega} (\sigma {\bf k} \rightarrow \sigma^{'} {\bf k^{'}},{\mathbf B})
= \frac{| f^{0} (\sigma {\bf k} \rightarrow \sigma^{'} {\bf k^{'}}) + f^{1} (\sigma {\bf k} \rightarrow \sigma^{'} {\bf k^{'}},{\mathbf B}) |^{2} }{(4 \pi)^{2}},
\end{equation}  
where $f^{0} (\sigma {\bf k} \rightarrow \sigma^{'} {\bf k^{'}}) $ is the scattering field amplitude independent of ${\mathbf B}$. Using Eq.~(7) to calculate $f^{1} (\sigma {\bf k} \rightarrow \sigma^{'} {\bf k^{'}})$
and inserting the resulting expression into
Eq.~(14) one obtains, after summing over outgoing polarization and after averaging
over incident polarization:

\begin{equation}
\label{eq:dif2}
\frac{d\sigma ^{1}}{d\Omega } = -2\frac{k\alpha (B)}{(4\pi )^{3}}\text{Re}\sum
_{ NN^{^{\prime }} LML^{\prime }} \text{Tr}\left( 
{\mathbf \Delta }_{{\mathbf k}}\cdot {\mathbf M}_{0}^{NN^{\prime }} \cdot {\mathbf \Delta }
_{{\mathbf k}^{\prime }} \cdot {\mathbf M}_{0}^{\ast ML^{\prime }}\cdot {\mathbf \Phi }\cdot 
{\mathbf M}_{0}^{\ast LM}\right) \exp \left[ -i {\mathbf k} \cdot {\mathbf r}_{NL}+i
{\mathbf k}^{\prime }\cdot {\mathbf r}_{N^{\prime }L^{\prime }}\right].
\end{equation}
Since we are interested in scattering quantities sensitive to chirality, we will again
perform an average of the scatterers positions over all rotations in Eq.~(15), as we have done for
the total MC cross-section. The calculation of the angularly 
averaged differential cross section linear in ${\mathbf B}$
$\left \langle \frac{d \sigma^{1}}{d \Omega} (\sigma {\bf k} \rightarrow \sigma^{'} {\bf k^{'}},{\mathbf B}) \right \rangle_{4 \pi}$
implies the construction of a rotationally invariant scalar from the
three vectors ${\mathbf B}$, ${\mathbf k}$ and ${\mathbf k}^{\prime}$. 
It can
be verified that the only possible form for 
$\left \langle \frac{d \sigma^{1}}{d \Omega} (\sigma {\bf k} \rightarrow \sigma^{'} {\bf k^{'}},{\mathbf B}) \right \rangle_{4 \pi}$
obeying parity and reciprocity relations is:
\begin{equation}
\left \langle \frac{d \sigma^{1}}{d \Omega} (\sigma {\bf k} \rightarrow \sigma^{'} {\bf k^{'}},{\mathbf B}) \right \rangle_{4 \pi}
= \cal{H} ( {\mathbf k} \cdot {\mathbf k}^{\prime} ) {\mathbf B} \cdot ( {\mathbf k} \times {\mathbf k}^{\prime} )  
+ \cal{C} ( {\mathbf k} \cdot {\mathbf k}^{\prime} ) {\mathbf B} \cdot ( {\mathbf k} + {\mathbf k}^{\prime} )  
\end{equation}
where $\cal{H} ( {\mathbf k} \cdot {\mathbf k}^{\prime} ) $ is a {\em scalar} representing the {\em Photonic Hall effect}~\cite{hall}
whereas $\cal{C} ( {\mathbf k} \cdot {\mathbf k}^{\prime} ) $ is a {\em pseudoscalar} associated with the MC effect.
We can thus explicitely obtain an expression for the function
$\cal{C} ( {\mathbf k} \cdot {\mathbf k}^{\prime} ) $ using Eq. (15) and by putting
$ {\widehat {\mathbf B}} = {\widehat {\mathbf k}} $ in Eq. (16).

	In the following, we will numerically calculate $g$ and 
$\left \langle \frac{d \sigma^{1}}{d \Omega} (\sigma {\bf k} \rightarrow \sigma^{'} {\bf k^{'}},{\mathbf B}) \right \rangle_{4 \pi}$.

%
%%%%%%%%%%%%%%%%%%%%
% Section 3:
%%%%%%%%%%%%%%%%%%%%
%
\section{Scattering chiral measures for the Twisted H}
\label{sec:twistedh}

	The ``twisted H", depicted in Fig.1, has the convenient property
that its chiral properties depend in a simple way on 
the angle $\gamma$ between its arms.
In Fig.~2 we have numerically calculated
the parameter $g$ in Eq.~(13) for 
four magneto-optically active pointlike scatterers placed at the vertices
of the twisted H as a function of the angle $\gamma$ for two different
values of the wavelength $\lambda$ of the incident light.
Each scatterer was set on resonance with the incident radiation, i.e., we set 
$\lambda = \lambda_{0}$. In addition,
$g$ was normalized by the normal extinction cross-section
of the system, which can, for the set parameters used here, be adequately 
approximated  by $\sigma_{0} = N \sigma_{r}$
(where $N=4$ and $\sigma_{r} = 3 \lambda^{2}_{0} / 2 \pi$ is the on-resonance extinction
cross-section for one scatterer) since we are in the independent scattering regime.
In Fig.~2 we see that $g$ exhibits an oscillatory behavior as 
a function of the angle $\gamma$ and {\em vanishes exactly 
at the configurations for which the H is achiral}. These configurations correspond
to the angles $\gamma = n \pi/2$, with $n$ integer. If
the value of the incident wavelength is modified, the dependence of
$g$ on $\gamma$ changes, as one can see in Fig.~2,
but $g$ still vanishes at the same angles. 
For different values of the wavelength $\lambda$, $g$ may also vanish at other values of $\gamma$,
but the zeros required by symmetry (at the achiral configurations) remain unchanged.
For comparison, we show by a dotted line the chiral parameter 
$\psi \sim \sin (2\gamma)$ for the twisted H proposed by Harris {\sl et al.}~\cite{harrisprl,harrisrmp,harrisferro} which,
for the set of values $kr = 2.0$ and $kd = 6.0$, nicely follows our optical parameter $g$.

	In Fig.~3, we have numerically calculated the differential MC cross section 
averaged over all solid rotations,
$\langle \frac{d \sigma^{1}}{d \Omega} (\sigma {\bf k} \rightarrow \sigma^{'} {\bf k^{'}}, {\mathbf B}) \rangle_{4 \pi}$,  
for the ``twisted H" as a function of the angle $\gamma$ and for the
particular case where ${\mathbf k} \perp {\mathbf B}$ and ${\mathbf k} \perp {\mathbf k}^{\prime}$.
We observe again its
oscillatory dependence on $\gamma$ and, like $g$, its cancelation for achiral
configurations. This means that the intensity of the light scattered by an assembly
of magneto-optically active particles under the influence of a magnetic field can also
be regarded as an optical manifestation of the
degree of chirality associated with the geometrical
configuration of these scatterers in space.

	In spite of its simplicity, the ``twisted H" is an illustrative example
of a chiral system. The results presented here could be experimentally verified
by means of light scattering measurements performed in samples 
composed by identical Faraday-active ``twisted H" molecules dispersed in a liquid.
In the single scattering regime this directly provides 
the optical parameters $g$ and
$\langle \frac{d \sigma^{1}}{d \Omega} (\sigma {\bf k} \rightarrow \sigma^{'} {\bf k^{'}}, {\mathbf B}) \rangle_{4 \pi}$.

	In the following, we will focus our attention on multiple scattering
of light by randomly distributed particles.

%
%%%%%%%%%%%%%%%%%%%%
% Section 4:
%%%%%%%%%%%%%%%%%%%%
%

\section{Probing the chirality of random scattering systems} 
\label{sec:random}
		
	A system composed of a large number of randomly distributed
particles will in general be chiral. Our purpose will be to
quantify the degree of chirality of this kind of system using
the scattering parameter $g$ defined in Eq.~(13). 
We have numerically calculated $g$ for 1000 different
random configurations of $N$ scatterers
distributed in a sphere of radius $R$ and volume $V$, 
and have studied the
behavior of $g$ as a function of $N$ for two distinct situations:
increasing $N$ upon keeping $V$ constant and
increasing $N$ upon keeping the density $\rho = N/V$ of the scatterers constant.

	In Fig.~4, we show a typical histogram for the values of $g$ for 1000
realizations of $N=10$ scatterers randomly distributed
in a sphere. The histogram shows a distribution
centered at the origin, as expected on the basis of the law
of large numbers. Since we have a large number
of realizations of the disorder, the 
mirror image of any configuration is equally probable, so that $\langle g \rangle = 0$.
We will consider the variance $\Delta g \equiv \sqrt{ \langle g^{2} \rangle -  \langle g \rangle^{2} }$
of $g$ as a
candidate to probe the typical degree of chirality of an arbitrary random configuration.

	In Fig. 5, we plot the values of $\Delta g$ as a function of the number
of particles distributed in a sphere with $\rho$ constant.
The variances $\Delta g$ were determined 
by taking the full width at half maximum of the Gaussian fit to
the histograms of $g$ for $1000$ different realizations of the disorder (see Fig. 4).
One relevant dimensionless parameter in the problem 
is the quantity $\zeta \equiv \frac{4 \pi \rho}{k^{3}}$,
which is essentially the number of particles per cubic wavelength. 
For the value of $\rho$ chosen in Fig. 5
we have $\zeta \approx 0.01$, which means that we are in the
so-called independent scattering regime and that 
only the first orders of scattering are relevant.
This allows us to normalize $\Delta g$ in Fig. 5 by $\sigma_{0} = N \sigma_{r}$.
As expected, $\Delta g$ vanishes for $N=2, 3$, since it
is impossible to generate a chiral configuration with only two or 
three particles.
In addition, we observe that
the normalized $\Delta g$ increases until it reaches, 
for roughly $N = 30$ particles in the sphere, an asymptotic value
of approximately $\frac{\Delta g}{\sigma_{0}} \approx 1.35$ x $10^{-4}$.
For comparison, we also show in Fig. 5 the behaviour of the
variance $\Delta \psi$ of the chiral parameter $\psi$
proposed by Harris {\sl et al.}~\cite{harrisprl,harrisrmp,harrisferro} as a function of $N$.
The comparison between the two curves in Fig. 5 clearly
reveals that the two chiral measures $g$ and $\psi$,
are, statistically speaking, {\em proportional}.
The observed relation of proportionality between 
$\Delta g$ and $\Delta \psi$ is given by:
\begin{equation}
\frac{\Delta g}{\sigma_{0}} \propto \frac{\Delta \psi}{R^{8}} \lambda^{3},
\end{equation} 
showing that we can establish a correspondence between the optical chiral
measure $g$ and the purely geometrical one $\psi/R^{8}$ by multipling the latter
by the optical quantity $\lambda^{3}$.
This demonstrates that
the correspondence between $g$ and $\psi$ is valid not only for the
``twisted H" case discussed in Sec. III, but applies more generally to random systems.

	In Fig. 6, $\Delta g$ and $\Delta \psi$ are shown as a function of $N$
if $V$ is kept constant, i.e., where $\rho$
was varied. In this case, $\zeta \approx 0.00025 N$.
We observe that both $\Delta g$ and $\Delta \psi$ 
increase linearly with the number of scatterers.
The comparison between $\Delta g$ and $\Delta \psi$ in Fig. 6
confirms, for $\rho$ constant, the relation of proportionality
between these two chiral parameters.
In order to understand the linear dependence of $\Delta g$ on $N$
we recall that we need at least four particles
to constitute a chiral system. This suggests to consider the 
chiral parameter $g$ as the sum of the
contributions of groups of four particles,
with random sign but with fixed absolute value $g_{4}$.
Since $\langle g \rangle = 0$ and since
$ \left( 
\begin{array}{c}
N \\ 
4
\end{array}
\right) $
distinct forms exist to group four particles together, 
we can estimate the normalized $\Delta g$ as: 
\begin{equation}
\label{eq:deltag}
\frac{\Delta g}{\sigma_{0}} = \frac{\sqrt{\langle g^{2} \rangle}}{N \sigma_{r}}
\sim \frac{1}{N}
\sqrt{ \left( 
\begin{array}{c}
N \\ 
4
\end{array}
\right) (g_{4})^{2} }.
\end{equation}
Taking the limit $N \rightarrow \infty$ and
noticing by the analysis of Eq.~(13) that $g_{4} \sim 1/(kR)^{3}$,
we can rewrite Eq.~(18) as:
\begin{equation}
\frac{\Delta g}{\sigma_{0}} \sim 
\frac{\sqrt{N(N-1)(N-2)(N-3)}}{N(kR)^{3}} \sim
\frac{N}{(kR)^{3}} + ... \sim \rho \lambda^{3} + ... \text{ \ \ \ }.
\end{equation}
This heuristic argument confirms the linear dependence of
$\Delta g$ on $N$ observed in Fig. 6 for large $N$.
However, for larger densities (typically for $\zeta \approx 1$) this
relation is lost. 

%
%%%%%%%%%%%%%%%%%%%%
% Section 8:
%%%%%%%%%%%%%%%%%%%%
%

\section{Summary and Conclusion}
\label{sec:conclusion}

In this paper we have investigated scattering
of light in magneto-chiral (MC) media. In order
to generate MC effects,
we have studied multiple light scattering from chiral
geometries of magneto-optically active pointlike scatterers. 
We have calculated, for the first time, the total and the differential
scattering cross-section of such media, showing that they 
are proportional to pseudoscalars quantities. We have concluded that 
light scattering in MC media
is sensitive to the degree of chirality exhibited by the geometrical
distribution of the scatterers in the space. This 
constitutes a new optical manifestation of chirality, 
in addition to the well-known optical rotatory power. We have introduced in Sec. II
a new parameter $g$ to measure or to quantify the degree of chirality of
an arbitrary configuration of particles subject to MC effects, whose construction is
based on the light scattered properties by such a configuration. As a genuine
MC parameter, $g$ is linear in the external magnetic field and independent
of the polarization, contrary to the optical rotatory power.
We have shown that $g$ is a pseudoscalar
and vanishes for configurations composed of two or three particles, as required for
any chiral measure.

	In Sec. III we have numerically calculated $g$
for scatterers placed on the vertices of one of the simplest chiral objects,
the ``twisted H". We have shown that $g$ vanishes when the H is achiral.
Furthermore, $g$ has the same sinusoidal behaviour 
$g \sim \sin(2 \gamma)$ as the
chiral order parameter $\psi$ defined by Harris
{\em et al.}~\cite{harrisprl,harrisrmp,harrisferro} for the ``twisted H".
In Sec. IV we have calculated $g$ for $N$ randomly distributed 
scatterers in a sphere of volume $V$ and density $\rho$, a system
which will, in general, be chiral. We have studied the behaviour of the variance $\Delta g$
as a function of $N$
for two distinct
cases: increasing $N$ upon keeping $\rho$ constant and increasing $N$ upon keeping
$V$ constant, i.e., increasing $\rho$. We have concluded that the variances $\Delta g$ and
$\Delta \psi$ are proportional.

%
%%%%%%%%%%%%%%%%%%%%
% Acknowledgements:
%%%%%%%%%%%%%%%%%%%%
%
\acknowledgements

	We gratefully acknowledge G. D\"{u}chs, G. Rikken, 
G. Wagni\`ere, M. Rusek and A. Orlowski for fruitful discussions.
This work was supported by the Polonium contract 03290 VK
of the French Ministry of Foreign Affairs.
One of us (F.A.P.) also wishes to thank
CNPq/Brazil for financial support.

%
%%%%%%%%%%%%%%%%%%%%%%%%%%%%%%%%%%%%%%%%%%%%%%%%%%%%
%%%%%%%%%%%%%%%%
%
%   R E F E R E N C E S:
%
%%%%%%%%%%%%%%%%%%%%%%%%%%%%%%%%%%%%%%%%%%%%%%%%%%%%
%%%%%%%%%%%%%%%%
%

%
%
%
%%%%%%%%%%%%%%%%%%%%%%%%%%%%%%%%%%%%%%%%%%%%%%%%%%%%
%%%%%%%%%%%%%%%%
%
%    F I G U R E -- C A P T I O N S:
%
%%%%%%%%%%%%%%%%%%%%%%%%%%%%%%%%%%%%%%%%%%%%%%%%%%%%
%%%%%%%%%%%%%%%%
%
\newpage
\begin{figure}
\caption{Four scatterers located at the vertices of
the simplest chiral geometry: the so-called ``twisted H".
The coordinates of the scatterers are: $[\pm \frac{r}{2} \cos(\frac{\gamma}{2}), \pm \frac{r}{2} \sin(\frac{\gamma}{2}), \pm \frac{d}{2}]$.}
\end{figure}
%[1]
\begin{figure}
\caption{The on-resonance optical chiral parameter $g$ plotted as a function of the
angle $\gamma$ between the ``twisted H" arms. 
The solid curve corresponds to the values $kr =2.0$ and $kd = 6.0$
(with $k$ the light wavenumber). The
dashed curve corresponds to the values $kr =3.0$ and $kd = 4.0$.
The dotted curve presents the chiral parameter $\psi \sim \sin(2 \gamma)$
introduced in Refs.~[19-21]. 
$g$ was normalized
by $\sigma_{0} = N \sigma_{r}$, where $N=4$ and $\sigma_{r} = 3 \lambda^{2}_{0} / 2 \pi$.}
\end{figure}
%[2]
\begin{figure}
\caption{The on-resonance magneto-chiral differential scattering cross-section averaged over
all solid rotations, $\langle \frac{d \sigma^{1}}{d \Omega} (\sigma {\bf k} \rightarrow \sigma^{'} {\bf k^{'}}, {\mathbf B}) \rangle_{4 \pi}$,
plotted as a function of $\gamma$ for the ``twisted H" with
$kr =3.0$ and $kd = 9.0$ and for 
${\mathbf k} \perp {\mathbf B}$ and ${\mathbf k} \perp {\mathbf k}^{\prime}$ (solid curve).
The dotted curve presents the chiral parameter $\psi \sim \sin(2 \gamma)$
introduced in Refs.~[19-21]. $\langle \frac{d \sigma^{1}}{d \Omega} (\sigma {\bf k} \rightarrow \sigma^{'} {\bf k^{'}}, {\mathbf B}) \rangle_{4 \pi}$
was normalized by $\sigma_{0}$.}
\end{figure}
%[3]
\begin{figure}
\caption{A typical histogram for the on-resonance values of $g$ associated
with $1000$ different realizations of the positions of $N = 10$ scatterers
randomly distributed in a sphere of density $\rho$ constant. The values of
$g$ were normalized by $\sigma_{0}$ and the parameter $\zeta$, defined in the text, equals $\zeta \approx 0.01$.
The dotted curve corresponds to a Gaussian fit to the data, whose 
full width at half maximum was used to determine the variance $\Delta g$.}
\end{figure}
%[4]
\begin{figure}
\caption{The variances $\Delta g$ and $\Delta \psi$ of the chiral parameters
$g$ (full circles and solid line) and $\psi$ (empty squares and dotted line),
 obtained from $1000$ different realizations of the disorder, 
as a function of the number $N$ of scatterers randomly distributed in a sphere of density constant
and with $\zeta \approx 0.01$ .
The values $g$ were calculated on-resonance and normalized by $\sigma_{0} = N \sigma_{r}$.
In order to show dimensionless quantities, the values of $\psi$ were multiplied
by $\lambda^{3}/R^{8}$ (where $\lambda$ is the light wavelength and $R$ is the radius of the sphere).
This reveals the proportionality relation 
$\frac{\Delta g}{\sigma_{0}} \propto \frac{\Delta \psi}{R^{8}} \lambda^{3}$
between $\Delta g$ and $\Delta \psi$.
To allow a better comparison between $\Delta g$ and $\Delta \psi$,
we have multiplied the values of $\Delta \psi$ by an appropriate constant numerical factor.
The lines are just a guide for the eyes.}
\end{figure}
%[5]
\begin{figure}
\caption{As in Figure 5, but now with $N$ scatterers distributed in a
sphere at constant volume and with $\zeta \approx 0.00025N$.}
\end{figure}

%
%
%%%%%%%%%%%%%%%%%%%%%%%%%%%%%%%%%%%%%%%%%%%%%%%%%%%%
%%%%%%%%%%%%%%%%
%
%  T A B L E S:
%
%%%%%%%%%%%%%%%%%%%%%%%%%%%%%%%%%%%%%%%%%%%%%%%%%%%%
%%%%%%%%%%%%%%%%
%
%
%
%
%%%%%%%%%%%%%%%%%%%%%%%%%%%%%%%%%%%%%%%%%%%%%%%%%%%%
%%%%%%%%%%%%%%%%
%
%                       End of Document
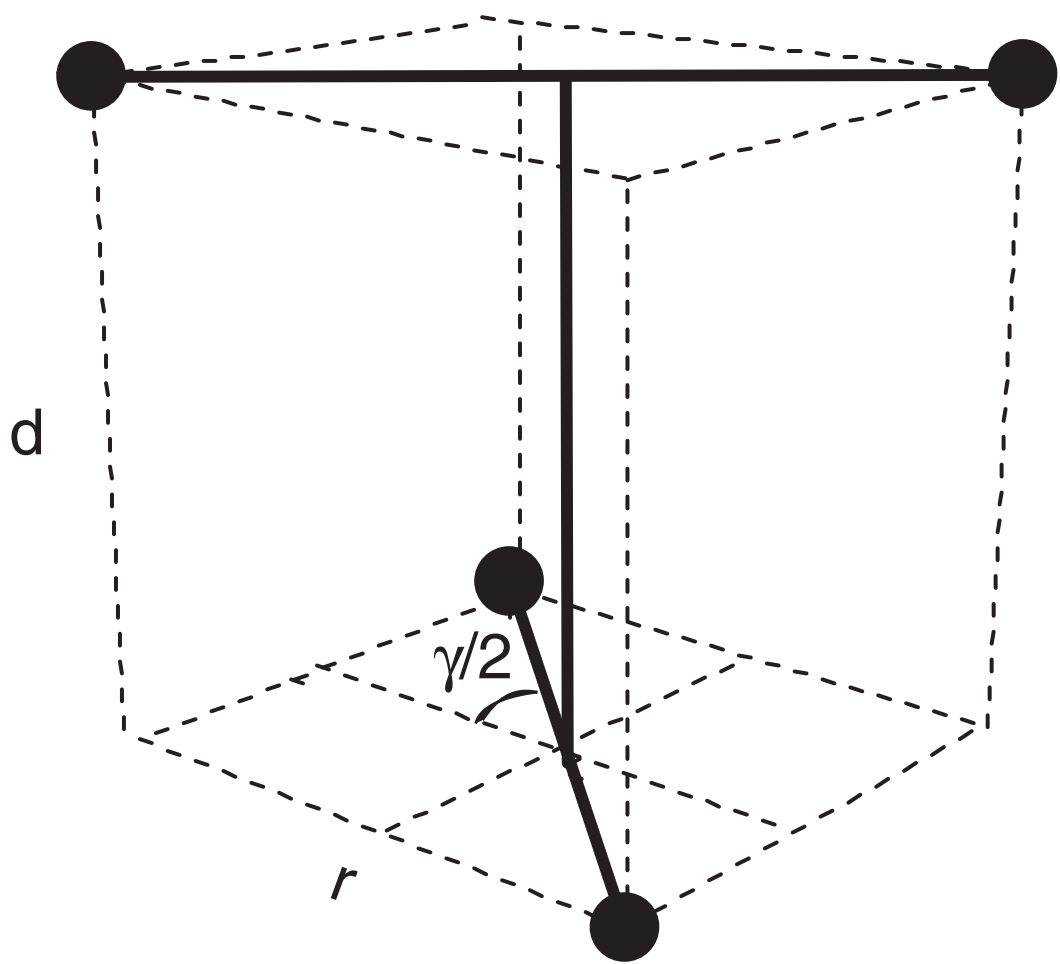
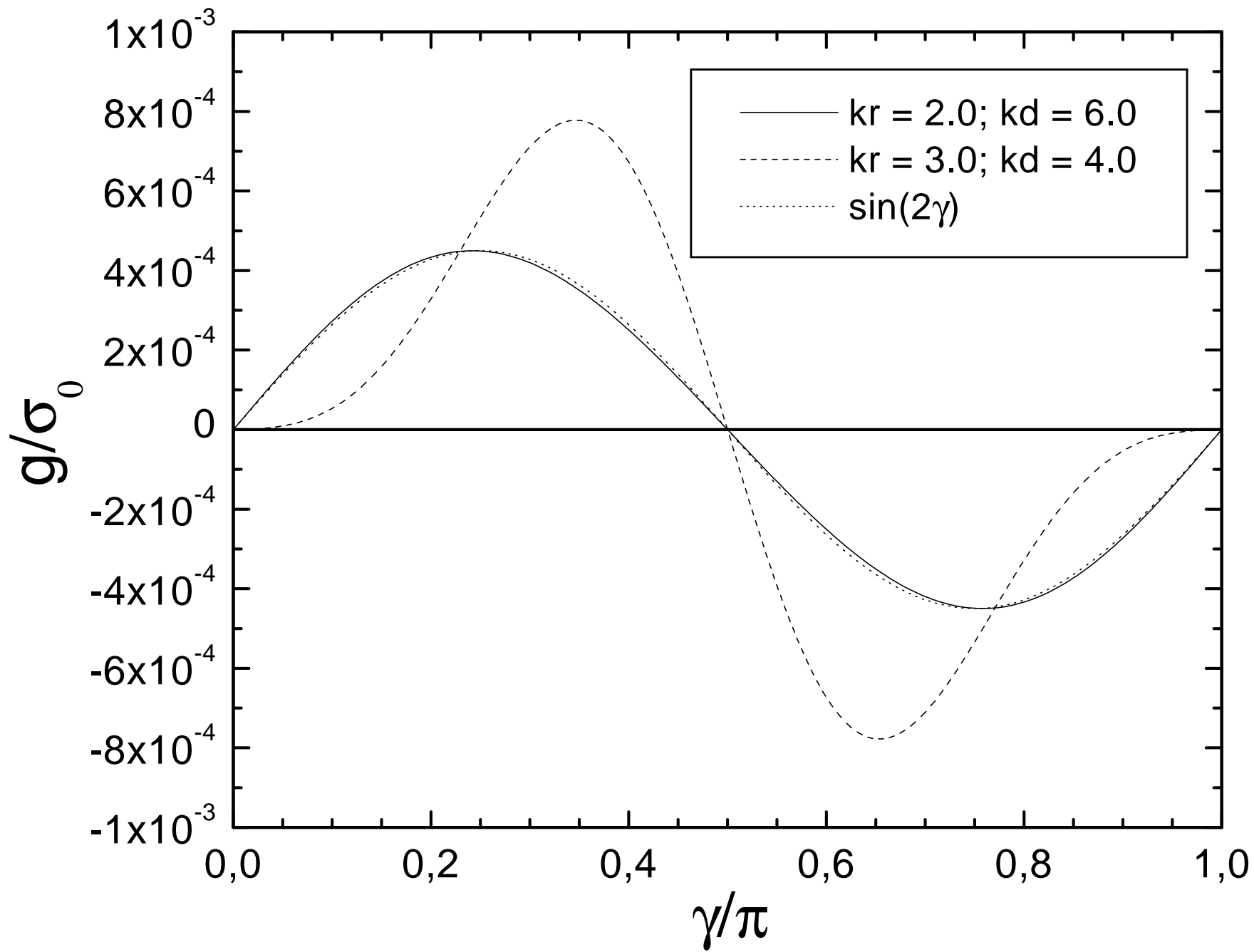
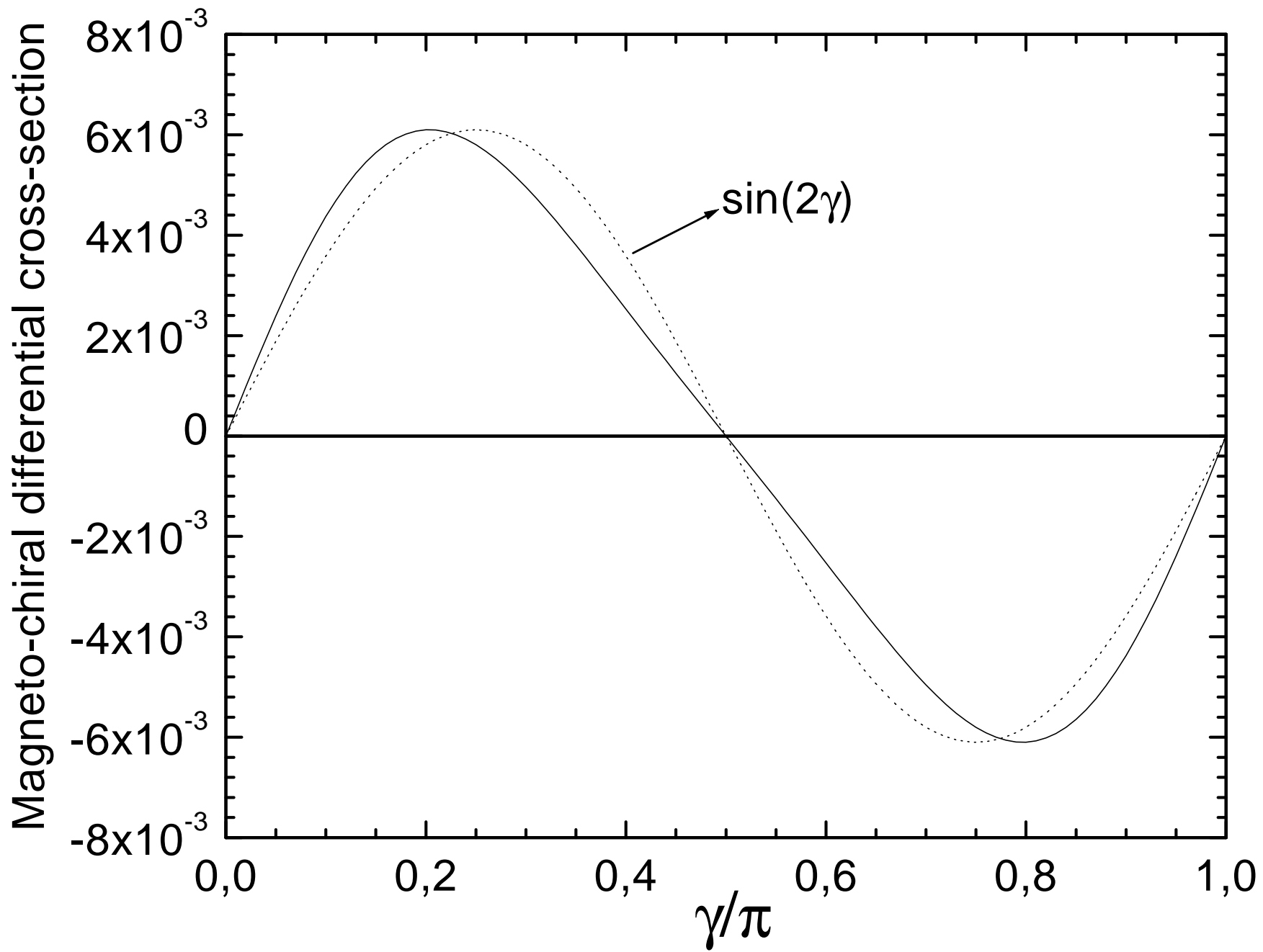
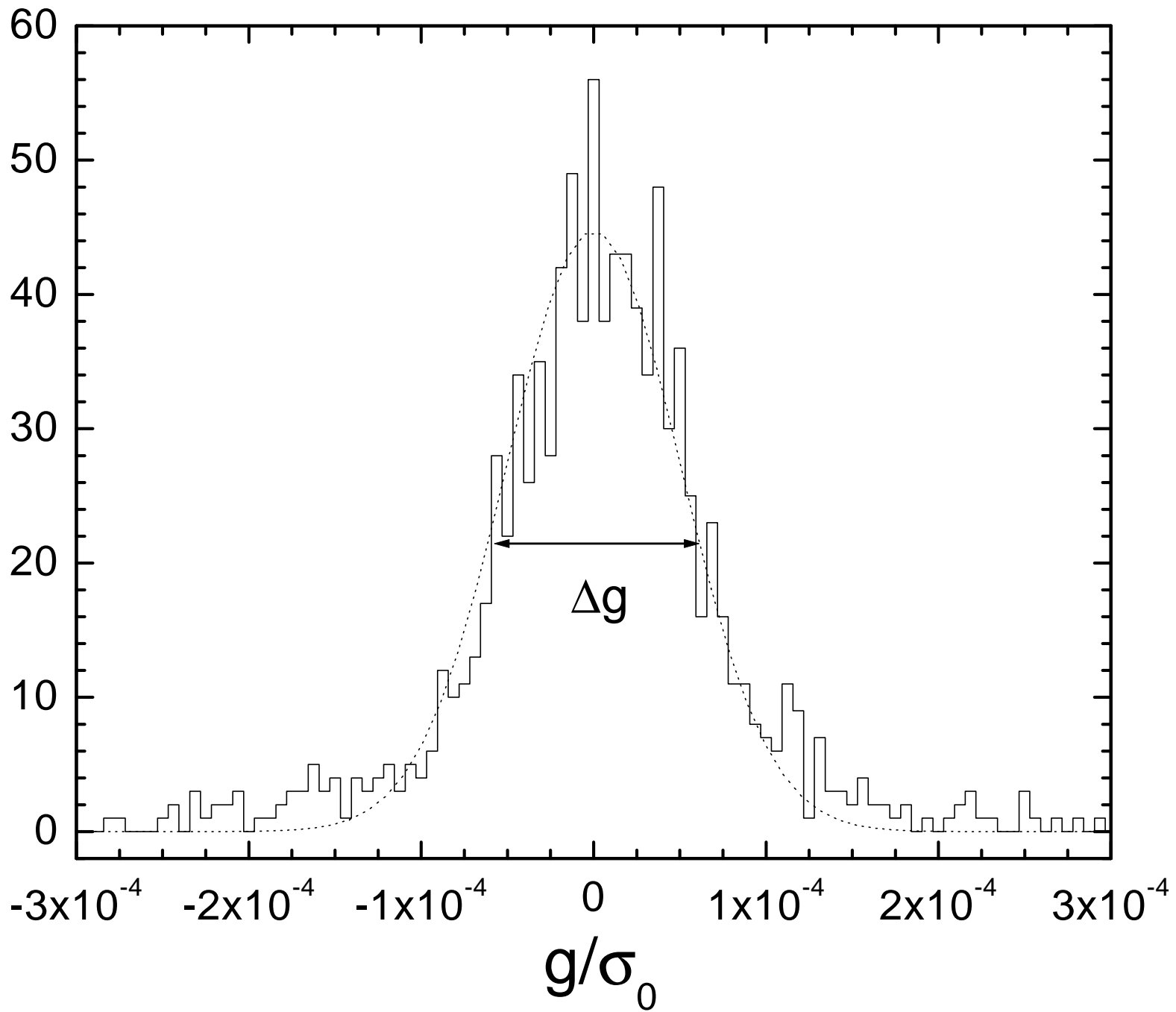
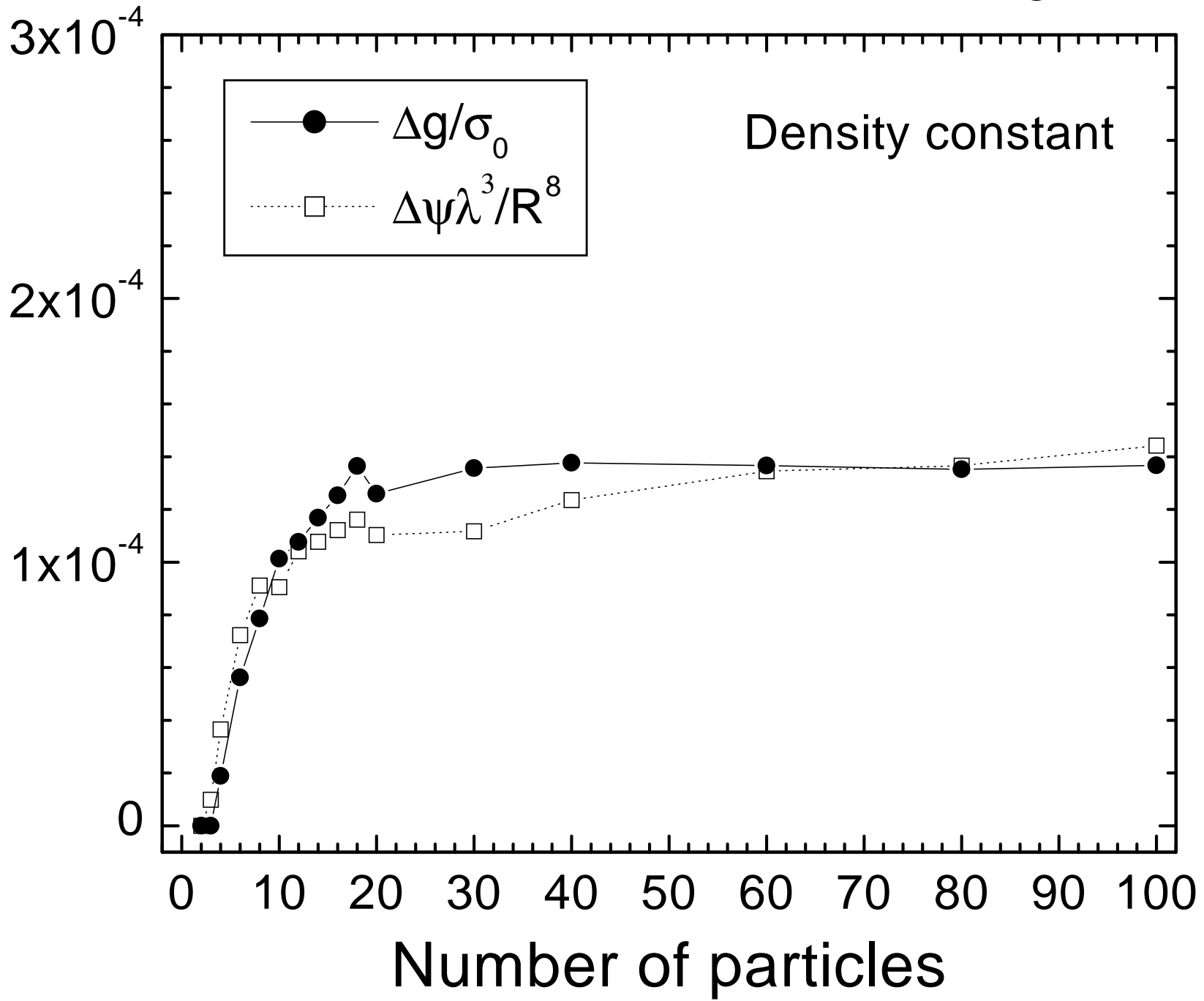
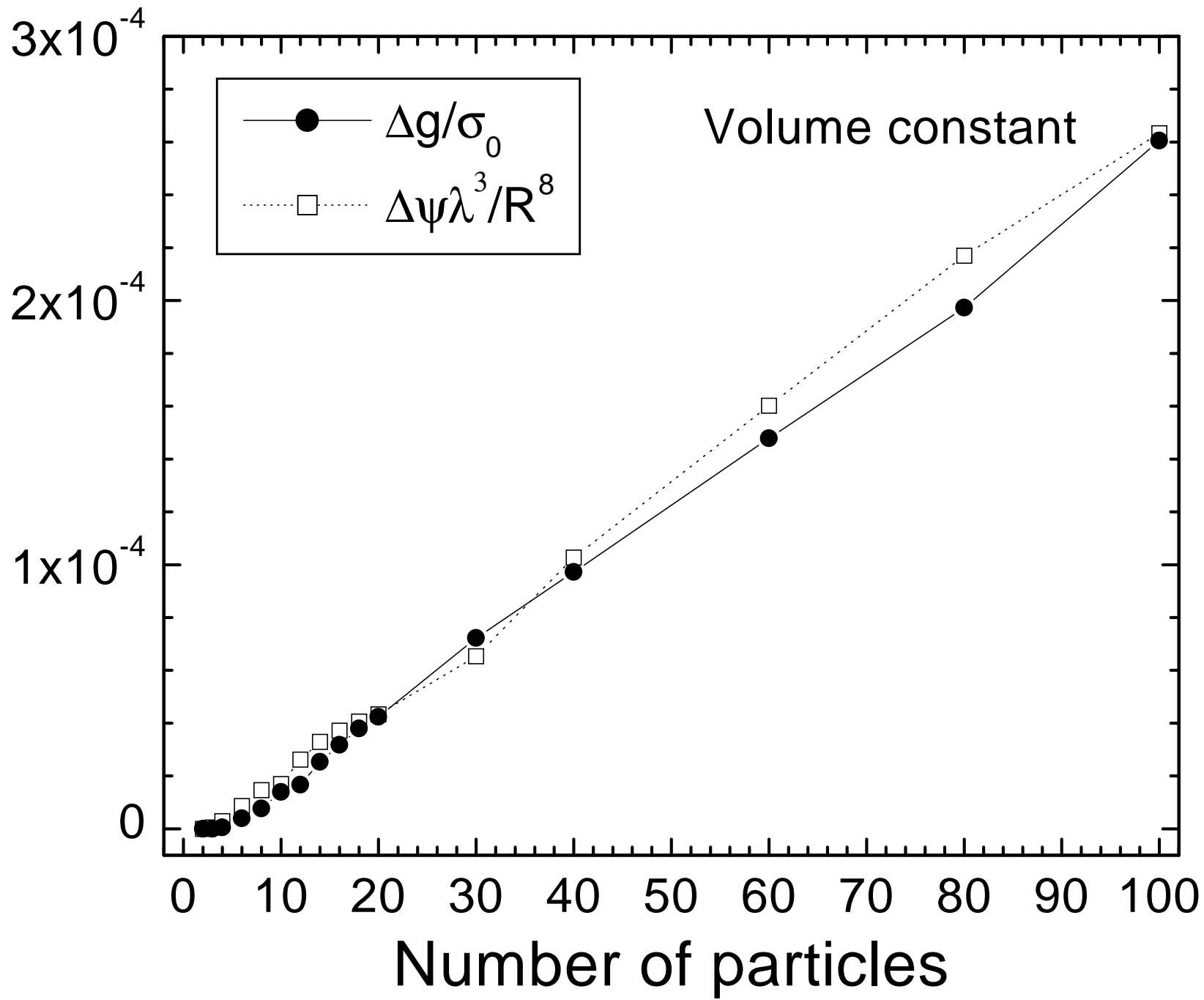
\end{document}